\documentclass[pra,twocolumn,floatfix]{revtex4}
\DeclareUnicodeCharacter{2009}{\,}
\usepackage{graphicx}
\usepackage{color}
\usepackage{amsmath}
\begin{document}

\title{Excitation spectrum of a bright solitary wave in a Bose-Einstein condensate
and its connection with the Higgs and the Goldstone modes}

\author{G. M. Kavoulakis$^{1,2,*}$}
\affiliation{$^1$Department of Mechanical Engineering, Hellenic Mediterranean University, 
P.O. Box 1939, GR-71004, Heraklion, Greece
\\
$^2$HMU Research Center, Institute of Emerging Technologies, GR-71004, Heraklion, Greece
\\
$^*$Corresponding author: G. M. Kavoulakis, email address: kavoulak@cs.hmu.gr}
\date{\today}

\begin{abstract}

We consider the problem of Bose-Einstein condensed atoms, which are confined 
in a (quasi) one-dimensional toroidal potential. We focus on the case of an 
effective attractive interaction between the atoms. The formation of a localized 
blob (i.e., a ``bright" solitary wave) for sufficiently strong interactions 
provides an example of spontaneous symmetry breaking. We evaluate analytically and 
numerically the excitation spectrum for both cases of a homogeneous and of a 
localized density distribution. We identify in the excitation spectrum the 
emergence of the analogous to the Goldstone and the Higgs modes, evaluating 
various relevant observables, gaining insight into these two fundamental modes 
of excitation.

\end{abstract}

\maketitle

\section{Introduction}

The concept of spontaneous symmetry breaking plays a fundamental role in various
physical systems. Quite generally, one may say that spontaneous symmetry breaking 
occurs when, while a Hamiltonian has some continuous symmetry, the solution breaks 
this symmetry. It is well-known that when a system undergoes a transition to a 
state where the symmetry is broken, the collective fluctuations in the phase and 
in the density of the corresponding order parameter \cite{Pekker, Shimano} give 
rise to the Goldstone \cite{Goldstone} and to the Higgs modes \cite{Higgs}, 
respectively.

Numerous physical systems exhibit these modes. While giving a complete list of
such systems is beyond the scope of this study, we will mention just a few
examples. Starting with the field of particle physics, the Higgs mode gives 
mass to the elementary particles \cite{Ryder, Cms, Atlas}. Experimental evidence 
for the existence of the Higgs mode has been seen in various systems, including 
superconductors \cite{Klein, Littlewood, Matsunaga, Gallais, Sherman}, 
antiferromagnets \cite{Ruegg} and ultracold atoms \cite{Bissbort, Endres, Leonard,
Behrle, Bayha, Dyke, Kell, Cabrera}; see also the theoretical studies \cite{Liu, 
G1, G2}. 

In the present study we demonstrate the emergence of these two modes in a rather 
simple physical system, namely that of an atomic Bose-Einstein condensate which 
is confined in a (quasi) one-dimensional toroidal potential, for effective attractive 
interactions between the atoms. Such traps have been realized experimentally, see, 
e.g., Refs.\,\cite{ring1, ring2, ring3, ring4, ring5, ring6, ring7, ring8, ring85, 
ring9, ring10}. Furthermore, both the strength and the sign of the effective potential
that describes the atom-atom interaction may be tuned with use of the so-called 
Feshbach resonances \cite{Fesh}. Under the conditions of quasi-one-dimensional motion 
that we consider, when the coupling $\gamma$ becomes smaller than some critical value 
$\gamma_c$, there is a phase transition from a homogeneous density distribution to 
a localized blob (i.e., a ``bright" solitary wave) \cite{Ueda0, Kavoulakis1, Ueda1, 
Ueda2}. Thus, we have an example of spontaneous symmetry breaking. 

The main result of this article is the excitation spectrum of the system for both 
cases where $\gamma$ is smaller and larger than $\gamma_c$. Our analytic results 
for the solution of the many-body problem are exact in the limit of large $N$, 
with $\gamma$ kept fixed, as is also seen from the comparison with the full 
numerical solutions that result from the diagonalization of the many-body 
Hamiltonian. 

In what follows below we start in Sec.\,II with the model that we adopt. Then, 
in Sec.\,III we demonstrate the emergence of the mechanism of spontaneous 
symmetry breaking in our problem. In Secs.\,IV and V we diagonalize analytically 
the Hamiltonian in the two cases of a homogeneous and an inhomogeneous density 
distribution, respectively. In Sec.\,VI we analyse the derived excitation 
spectrum, and we compare it with the solutions that result from the numerical 
diagonalization of the many-body Hamiltonian. Also, we discuss its relevance 
with the Higgs and the Goldstone modes. Finally we comment on the experimental 
relevance of our results. In the last section, Sec.\,VII, we provide an 
overview of the basic results of our study.

\section{Model}

The physical system that we have in mind is that of Bose-Einstein condensed atoms 
which are confined in a toroidal potential, with a strong confinement in the 
transverse direction, which makes their motion (quasi) one-dimensional. We thus 
work with the basis of the single-particle eigenstates of a purely one-dimensional 
potential under periodic boundary conditions, i.e., $\phi_m(\theta) = 
e^{i m \theta}/\sqrt{2 \pi R}$ with eigenvalues $e_m = \hbar^2 m^2/(2 M R^2)$. 
Here $m$ is the quantum number that corresponds to the single-particle angular momentum, 
$\theta$ is the azimuthal angle, $M$ is the atom mass, and $R$ is the ``mean" radius of 
the torus. We model the atom-atom interactions as a contact interaction. The matrix
element $U_0$ for s-wave, elastic atom-atom collisions is given by $U_0 = 2 \hbar^2 
a/(M R S)$. Here $a$ is the scattering length for atom-atom collisions, and $S$ is the 
cross section of the torus, with $\sqrt{S} \ll R$. The Hamiltonian of the system is, 
thus, 
\begin{eqnarray}
   {\hat H} = \frac {\hbar^2} {2 M R^2} \sum {\hat c}_m^{\dagger} {\hat c}_m m^2 
   + \frac {U_0} {2} \sum_{m,n,l,k} {\hat c}_m^{\dagger} 
   {\hat c}_n^{\dagger} {\hat c}_l {\hat c}_k \, \delta_{m+n,l+k}.
\label{mbh}
\nonumber \\
\end{eqnarray}
Here ${\hat c}_m^{\dagger} ({\hat c}_m)$ is the creation (annihilation) operator 
of a particle with angular momentum equal to $m$. 

We introduce the dimensionless parameter $\gamma$, where $\gamma/2$ is the ratio 
between the interaction energy per particle of the homogeneous phase, $N \hbar^2 a/
(M R S)$, and the kinetic energy $e_1 = \hbar^2/(2 M R^2)$, i.e., $\gamma = 4 N a R/S$, 
with $N$ being the total number of atoms. From now on $\hbar$, $2M$, and $R$ are set 
equal to unity. For $\gamma < 0$  we have (effective) attractive interactions, while 
for $\gamma > 0$  we have (effective) repulsive interactions. As we explain in detail 
in Sec.\,III, within the mean-field approximation, when $\gamma > \gamma_c = -1/2$ the 
density is homogeneous, whereas when $\gamma < \gamma_c = -1/2$ there is a (continuous) 
phase transition to a localized density distribution. Clearly, as $\gamma$ decreases 
below the critical value $\gamma_c = -1/2$, the density distribution becomes more and 
more narrow. 

In the results that follow below we work with the Hamiltonian of Eq.\,(\ref{mbh}), 
where we restrict ourselves to the subspace of single-particle states with $m = -1, 
0$ and 1. Within the mean-field approximation (similar arguments also
hold for the many-body state) one may expand the order parameter $\Psi(\theta)$ in 
the single-particle states $\phi_m$, i.e., 
\begin{equation}
   \Psi(\theta) = \sum_m d_m \phi_m(\theta).
\label{expan}
\end{equation}
In the homogeneous phase the order parameter is given trivially by $\phi_0(\theta)$. 
The crucial observation here is that close to the transition to the localized phase, 
i.e., for $\gamma \lesssim \gamma_c = -1/2$, the single-particle states with the dominant 
contribution to the order parameter are the ones with $m=-1$ and $+1$ (in addition, of 
course, to the state with $m = 0$). This is due to the fact that the single-particle 
energy $e_m$ increases quadratically with the quantum number of the single-particle 
angular momentum $m$, i.e., $e_m \propto m^2$. An order-of-magnitude estimate
which is based on the comparison between the kinetic and the interaction energies 
implies that the states with a significant contribution are roughly the ones with 
$|m| \le \sqrt{|\gamma|}$. Actually, close to the transition and in the regime of 
a localized density distribution, $\gamma \lesssim \gamma_c = -1/2$, one may develop 
a power-series expansion for the amplitudes $d_m$ in the small parameter 
$\gamma + \gamma_c$. As a result, the parameters $d_m$ are more and more suppressed
with increasing $|m|$, for $|m| > 1$ \cite{Kavoulakis1}.

While we use the mean-field approximation in Sec.\,III, in the rest of the paper 
we adopt the method of diagonalization of the many-body Hamiltonian. Within the 
mean-field approximation the many-body state is assumed to have a product form. 
In this approach one makes the implicit assumption of $N \to \infty$ and $L \to 
\infty$, where $L$ is the total angular momentum, with $L/N$ and $\gamma$ fixed. 
Within the diagonalization of the many-body Hamiltonian we work with a finite $N$ 
and $L$, considering the eigenstates of the operators of the atom number and of the 
total angular momentum, and diagonalize the resulting Hamiltonian matrix, both 
analytically and numerically. We stress that the single-particle density distribution 
that results from the eigenstates of the Hamiltonian that we evaluate is always 
axially-symmetric. This is due to the fact that we work with eigenstates of the 
operator of the angular momentum. 

\begin{figure}
\includegraphics[width=8cm,height=7cm,angle=-0]{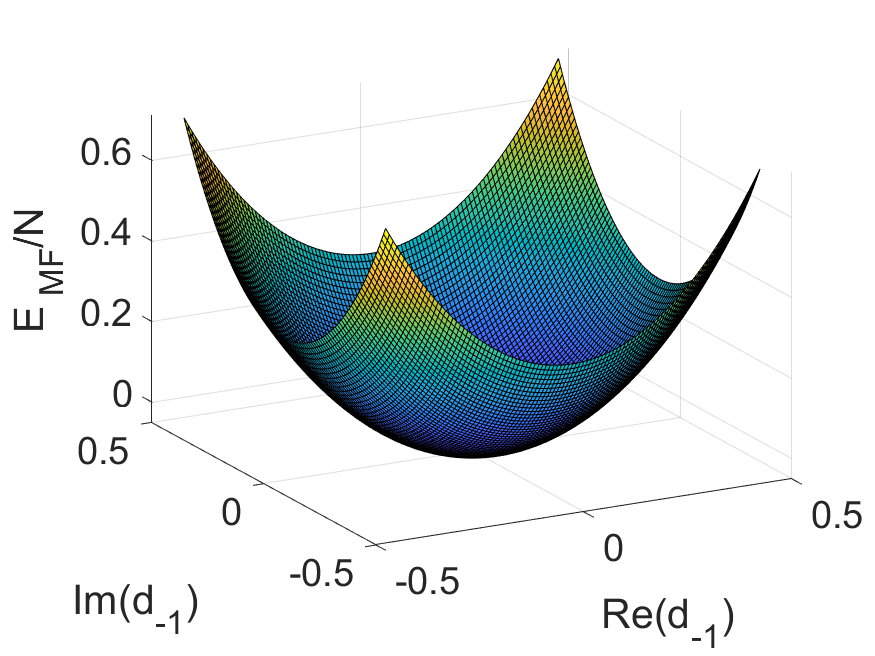}
\includegraphics[width=8cm,height=7cm,angle=-0]{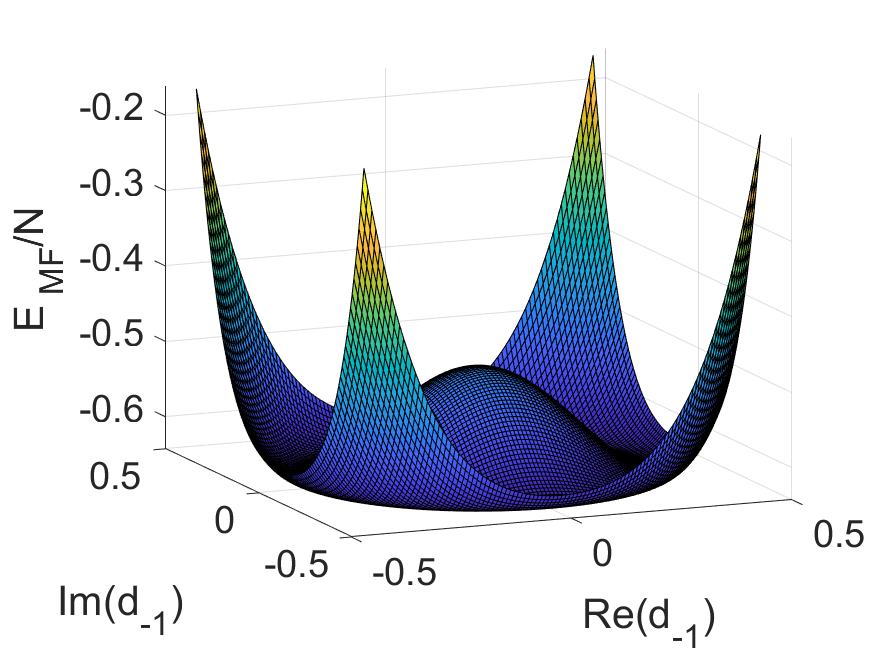}
\caption{The energy $E_{\rm MF}/N$ of Eq.\,(\ref{emffg}) (in units of $e_1$) for $L/N = 0$ 
and for two values of $\gamma$, as function of Re$(d_{-1})$ and Im$(d_{-1})$. In the upper 
plot $\gamma = -0.1 > \gamma_c$ and the minimum of the potential is at the center, where 
$d_{-1} = d_1 = 0$. As a result, the density distribution of the atoms is homogeneous. 
In the lower plot $\gamma = -1 < \gamma_c$, $|d_{-1}| = |d_1| \neq 0$, and we have the 
formation of a localized blob (i.e., a ``bright" solitary wave).}
\end{figure}

\begin{figure}
\includegraphics[width=8cm,height=7cm,angle=-0]{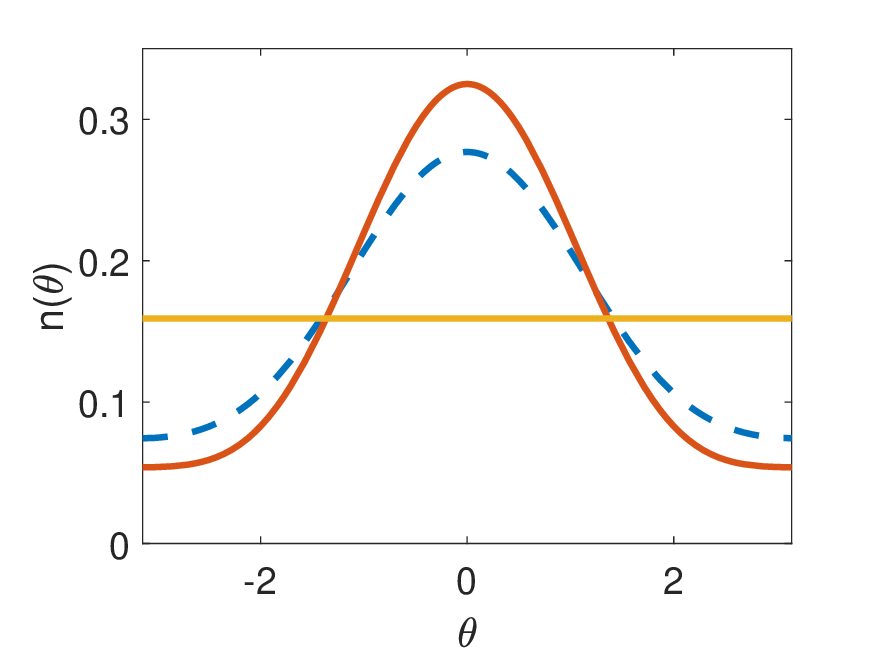}
\caption{The density $n(\theta) = |\Psi(\theta)|^2$ of Eq.\,(\ref{densityy}) (in 
units of $1/R$) for $L/N = 0$ and for three values of $\gamma > -1/2$ (straight,
horizontal line), $\gamma = -0.55$ (dashed curve) and $\gamma = -0.6$ (solid 
curve). Here $\varphi_c = 0$.}
\end{figure}

\section{Spontaneous symmetry breaking}

Before we proceed, it is instructive to demonstrate the mechanism of spontaneous 
symmetry breaking via an explicit calculation. Only in this section we thus work 
within the mean-field approximation. 

Let us consider the order parameter of Eq.\,(\ref{expan}), keeping only the states
with $m = 0$ and $m = \pm 1$ \cite{Kavoulakis1},
\begin{equation}
   \Psi(\theta) = d_{-1} \phi_{-1}(\theta) + d_0 \phi_0(\theta) + d_1 \phi_1(\theta). 
   \label{trial}  
\end{equation}
The normalization condition implies that $|d_{-1}|^2 + |d_0|^2 + |d_1|^2 
= 1$. Since we work with a fixed angular momentum, we impose the additional constraint
$|d_1|^2 - |d_{-1}|^2 = L/N$.

The expectation value of the energy per particle is given by
\begin{eqnarray}
   \frac {E_{\rm MF}} N =  
   \int_{-\pi R}^{\pi R} \left| \frac {\partial \Psi} {\partial \theta} \right|^2 \, dx
   +  2 \pi R \frac {\gamma} {2} \int_{-\pi R}^{\pi R} \left| \Psi \right|^4 \, dx.
\label{mbhhh}
\end{eqnarray} 
From Eqs.\,(\ref{trial}) and (\ref{mbhhh}) we find that 
\begin{eqnarray}
  \frac {E_{MF}} N = |d_{-1}|^2 + |d_1|^2 + 
  \nonumber \\
  +\frac {\gamma} 2 [|d_{-1}|^4 + |d_0|^4 + |d_1|^4 + 4 |d_{-1}|^2 |d_0|^2 
  \nonumber \\
  + 4 |d_{-1}|^2 |d_1|^2 + 4 |d_0|^2 |d_1|^2 + 2 |d_0|^2 (d_{-1} d_1 + d_{-1}^* d_1^*) ].
\label{emfff}
\end{eqnarray}
Since the overall phase of the order parameter is arbitrary, we assumed in 
Eq.\,(\ref{trial}) that $d_0$ is real. The two constraints of normalization and 
fixed angular momentum [see the two equations below Eq.\,(\ref{trial})] 
allow us to eliminate $d_0$ and $d_1$ and therefore express the energy in terms of 
$d_{-1}$, 
\begin{eqnarray}
  \frac {E_{MF}} N = \frac L N 
  + \frac {\gamma} 2 \left(1 + 2 \frac L N - 2 \frac {L^2} {N^2} \right) +
\nonumber \\  + 2 (1 + \gamma) |d_{-1}|^2 - 3 \gamma \frac L N |d_{-1}|^2 - 3 \gamma |d_{-1}|^4 + 
   \nonumber \\
   + 2 \gamma (1 - \frac L N - 2 |d_{-1}|^2) \sqrt{|d_{-1}|^2 (|{d_{-1}|^2 + \frac L N})}.
\label{emf}
\end{eqnarray}

For the special case $L = 0$, $|d_1| = |d_{-1}|$ and Eq.\,(\ref{emf}) takes the form
\begin{eqnarray}
  \frac {E_{MF}} N - \frac {\gamma} 2 =  2 (1 + 2 \gamma) |d_{-1}|^2 - 7 \gamma |d_{-1}|^4.
\label{emffg}
\end{eqnarray}
From this equation it follows that for $\gamma > -1/2$ the value of $d_{-1}$ that minimizes
the energy is zero, while for $\gamma < -1/2$, $|d_{-1}|^2 = (1 + 2 \gamma)/(7 \gamma)$. 

Indeed, in Fig.\,1 we plot the energy per particle $E_{MF}/N$ of Eq.\,(\ref{emffg}) as 
function of Re$(d_{-1})$ and Im$(d_{-1})$ for two values of $\gamma$, i.e., $\gamma = 
-0.1$ (higher, where the system is in the homogeneous phase) and $\gamma = -1$ (lower, 
where the system is in the inhomogeneous phase). As expected, we observe that for 
$\gamma =-0.1$ the minimum occurs at the center, i.e., $d_{-1} = 0$ and the density 
that corresponds to $\Psi$ is homogeneous. On the other hand, for $\gamma = -1$, we 
have the formation of a ``Mexican-hat" potential and $d_{-1} \neq 0$, i.e., the 
density becomes inhomogeneous. 

If $d_{\pm 1} = |d_{\pm 1}| e^{i \varphi_{\pm 1}}$, the minimization of the energy implies 
that $\varphi_{+1} = - \varphi_{-1} \equiv \varphi_c$.  For $\gamma < \gamma_c$, the density 
is given by
\begin{eqnarray}
  n(\theta) 
  = |\Psi(\theta)|^2 = \frac 1 {2 \pi} \left( 1 + 4 |d_0| |d_{-1}| \cos(\theta - \varphi_c) 
  +
\right. \nonumber \\ \left.
      + 2 |d_1|^2 \cos[2(\theta - \varphi_c)] \right).
      \label{densityy}
\end{eqnarray}
In Fig.\,2 we show the density $n(\theta) = |\Psi(\theta)|^2$, Eq.\,(\ref{densityy}), for 
three values of $\gamma$ and for $\varphi_c = 0$. The horizontal line corresponds to any 
value $\gamma > -1/2$ (i.e., the homogeneous phase), the dashed curve to $\gamma = -0.55$, 
and the solid curve to $\gamma = -0.6$.

We see that $n(\theta)$ depends on $\theta - \varphi_c$, while its maximum occurs at 
$\theta = \varphi_c$. Also, the energy is independent of the value of $\varphi_c$. In more
physical terms, there is a degeneracy, since the energy of the system is independent of the 
location of the blob. Therefore, we have an example of spontaneous symmetry breaking, where, 
although the Hamiltonian is axially symmetric, the solution breaks this symmetry.

\section{Symmetry-preserving solutions}

Let us start with the case $-1/2 < \gamma < 0$, where the system is in the homogeneous
phase. The analysis that we present below refers to the limit where the 
number of atoms $N$ is ``large", while $L$ is of order unity.

Working with the Fock states \cite{absl}
\begin{equation}
  | k \rangle = | (-1)^k, 0^{N - L - 2k}, (+1)^{k + L} \rangle, 
\label{Fock}
\end{equation}
we express the many-body states as
\begin{equation}
   |\Psi(L) \rangle = \sum_k f_k \, | k \rangle. 
   \label{mbs}
\end{equation}
We stress that these states are angular momentum eigenstates (we analyse this further
below). Let us now evaluate the matrix elements $H_{k,k'} = \langle k | 
{\hat H} | k' \rangle$. Starting with the diagonal ones,
\begin{eqnarray}
  H_{k, k} = 2k + L + 
  \nonumber \\
  + \frac {U_0} {4 \pi} [N (N-1) + 2 N L - 2 L^2 + 4 k N - 6 k L - 6 k^2].
\nonumber \\
\end{eqnarray}
Turning to the off-diagonal, 
\begin{eqnarray}
  H_{k, k+1} = 
  \nonumber \\
  = \frac {U_0} {2 \pi} \sqrt{ (N - L - 2 k) (N - L - 2 k -1) (k+1) (k + L +1)},
\nonumber \\
\end{eqnarray}
i.e., the Hamiltonian matrix is tridiagonal. For large values of $N$ we get that
\begin{eqnarray}
  H_{k,k} = L + \frac {\gamma} 2 \left[ (N-1) + 2 L - 2 \frac {L^2} N \right] + 
  \nonumber \\
  + 2 (1 + \gamma) k - 3 \gamma k \frac L N  - 3 \gamma \frac {k^2} N,
\end{eqnarray}
while
\begin{eqnarray}
 H_{k, k+1} &\approx& \frac {U_0} {2 \pi} (N - 2 k - L) \sqrt{k (k+L)} =
 \nonumber \\
 &=& \gamma N \left(1 - 2 \frac k N - \frac L N \right) \sqrt{\frac k N \frac {k+L} N}.
\end{eqnarray}
As seen in Fig.\,3, the values of $k$ with a significant contribution 
to the many-body state are of order unity. As mentioned before, $L$ is assumed to be 
of order unity, while $N$ is assumed to be ``large". Therefore, one may ignore the 
terms which are of order $k L/N$ and $k^2/N$, and as a result,
\begin{eqnarray}
  H_{k,k} - \frac {\gamma} 2 \left[(N - 1)- 2 \frac {L^2} N \right] = (1 + \gamma) (L + 2k).  
\end{eqnarray}
Also,
\begin{eqnarray}
  H_{k, k+1} = \gamma \sqrt{(k+L+1) (k+1)}.
\end{eqnarray}
From the last two equations it follows that the Hamiltonian may be written in the form
\begin{eqnarray}
  H - \frac {\gamma} 2 \left[(N - 1)- 2 \frac {L^2} N \right] = 
  \nonumber \\
  = (1 + \gamma) ({\hat c}_{-1}^{\dagger} {\hat c}_{-1} + {\hat c}_{1}^{\dagger} {\hat c}_{1}) 
  + \gamma ({\hat c}_{-1}^{\dagger} {\hat c}_{1}^{\dagger} + {\hat c}_{-1} {\hat c}_{1}). 
\end{eqnarray}
This Hamiltonian may be diagonalized via a Bogoliubov transformation, see, e.g., 
\cite{Kavoulakis1},
\begin{eqnarray}
   {\hat b} &=& \lambda_1 {\hat c}_{-1}^{\dagger} + \lambda_2 {\hat c}_1
   \nonumber \\
   {\hat d} &=& \lambda_2 {\hat c}_{-1} + \lambda_1 {\hat c}_1^{\dagger}.
\end{eqnarray}
Following the usual procedure, we get that the energy spectrum is given by
\begin{eqnarray}
  E_{n_b, n_d}(L) - \frac {\gamma} 2 \left[(N - 1)- 2 \frac {L^2} N \right] =
 \nonumber \\
  = - (1 + \gamma)
 + \sqrt{2 \gamma + 1} (1 + n_b + n_d),
\label{ess2}
\end{eqnarray} 
where $n_b$ and $n_d$ are the eigenvalues of the number operators ${\hat b}^{\dagger}
{\hat b}$ and ${\hat d}^{\dagger} {\hat d}$, respectively. For small $\gamma$ the above 
equation gives (ignoring terms of order $1/N$)
\begin{eqnarray}
  E_{n_b, n_d}(L) - \frac {\gamma} 2 (N - 1) = (1 + \gamma) (n_b + n_d).
\label{ess22}
\end{eqnarray}
Furthermore, ${\hat b}^{\dagger} {\hat b} - {\hat d}^{\dagger} d = c_1^{\dagger} c_1 
- c_{-1}^{\dagger} c_{-1} =  L$. Therefore, $n_b - n_d = L$ and as a result the energy 
spectrum is given by 
\begin{eqnarray}
  E_n(L) - N e_{\rm hom}(\gamma) = (2 n + L + 1) \, \omega_{\rm hom} - \gamma \frac {L^2} N.
\label{excsphom}
\end{eqnarray}
From the above equation it follows that
\begin{equation}
  \Delta E_n(L) = E_n(L) - E_0(0) = (2 n + L) \, \omega_{\rm hom} - \gamma \frac {L^2} N.
\label{excsphom1}
\end{equation}
Here $n = n_d = 0, 1, 2, \dots$ and also
\begin{equation}
  e_{\rm hom} =  \frac {\gamma} 2 - \frac {1 + \gamma} N + \frac {\sqrt{2 \gamma + 1}} N, 
\end{equation}
while
\begin{equation}
 \omega_{\rm hom} = \sqrt{2 \gamma + 1}. 
 \label{quhom}
\end{equation}
There are three important observations regarding Eq.\,(\ref{excsphom}): (i) There 
is a term which is linear in $L$, as expected, due to phonon excitation. (This is 
due to the fact that the single-particle density is homogeneous in this case). This 
linear term is still present, even if one includes more single-particle states, 
i.e., it is present independently of the imposed truncation. Furthermore, the 
coefficient of the linear term may be identified as the speed of sound $c$, i.e., 
$c = \omega_{\rm hom} = \sqrt{2 \gamma + 1}$. In the limit of small $\gamma$, 
$c \approx 1 + \gamma$. (ii) Apart from the last term on the right, which is of 
order $L^2/N$, the energy levels associated with both quantum numbers $n$ and 
$L$ are equally spaced and their energy difference is of order $\omega_{\rm hom}$, 
i.e., of order unity. (iii) The energy quantum $\omega_{\rm hom}$ vanishes for 
$\gamma = -1/2$ and is an increasing function of $\gamma$.

Regarding the amplitudes $f_k$ in Eq.\,(\ref{mbs}), these decay rapidly with $k$, 
as seen in Fig.\,3. This rapid decay has to do with the fact that the state is not 
fragmented (for $\gamma > -1/2$), i.e., $\langle k \rangle$ is of order unity 
\cite{frag}. In this figure we plot $f_{k}^2$ that we derive from the diagonalization
of the Hamiltonian for $\gamma = -0.1$, $N = 1000$, and $L = 0$. We also plot the
analytic result $f_k \approx (|\gamma|/2)^k$, which is valid for sufficiently small 
values of $|\gamma| \ll 1$. This result follows from the eigenvalue equation -- see
Eq.\,(\ref{eigeq}) below -- in the limit of small values of $\gamma$.

\begin{figure}
\includegraphics[width=8cm,height=7cm,angle=-0]{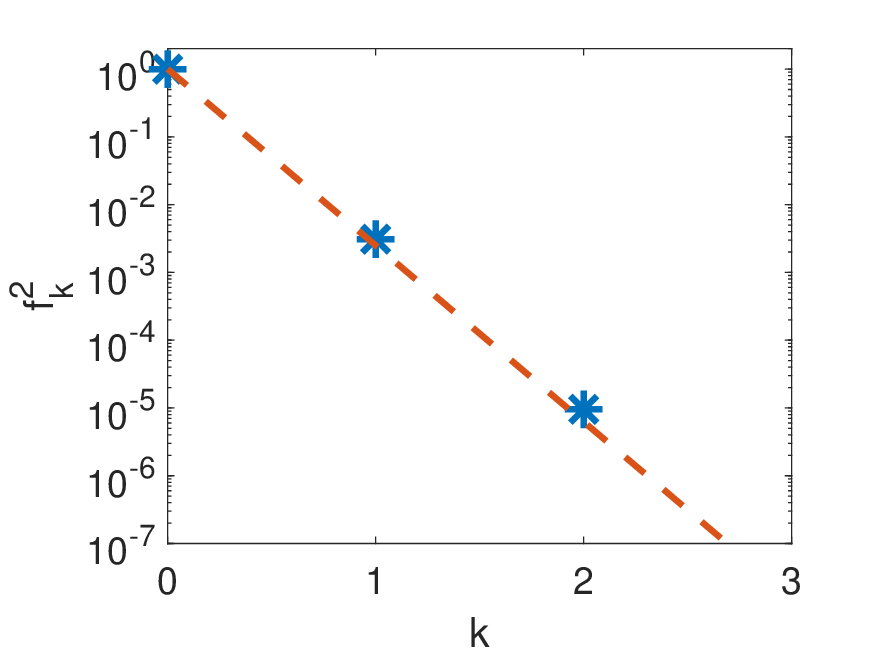}
\caption{Blue stars: The amplitudes $f_{k}^2$, Eq.\,(\ref{mbs}), for
$k = 0, 1$, and 2 (on a logarithmic $y$ scale), for $\gamma = -0.1$, which result 
from the numerical diagonalization of the many-body Hamiltonian, for $N = 1000$ 
and $L = 0$. Dashed curve: The analytic expression $f_k^2 = (|\gamma|/2)^{2k}$.}
\end{figure}

\section{Symmetry-breaking solutions}

For $\gamma < -1/2$ it is well-known that -- within the mean-field approximation -- 
a bright solitary wave forms, as we saw also in Sec.\,III. Starting with the amplitudes 
$f_k$, here we have a very different behaviour as compared to the previous section. 
As seen in Fig.\,4, the amplitudes peak around some $k_0$ (which is of order $N$). 
This reflects the fact that all three single-particle states are macroscopically 
occupied, and as a result the many-body state is fragmented \cite{frag}.

\begin{figure}
\includegraphics[width=8cm,height=7cm,angle=-0]{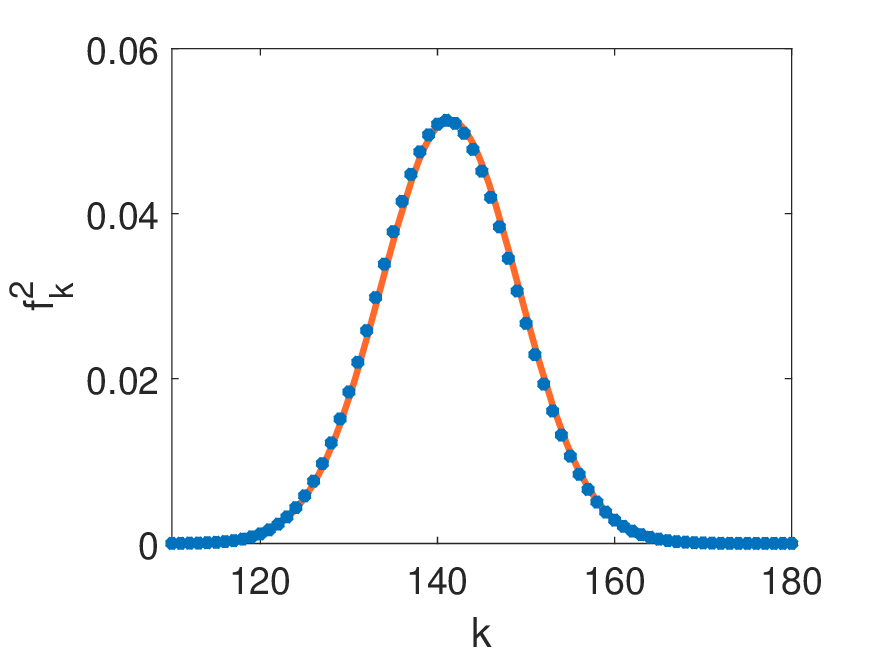}
\caption{Blue dots: The amplitudes $f_{k}^2$, Eq.\,(\ref{mbs}), for $k = 0, 
1, 2, \dots, 180$ which result from the numerical diagonalization of the many-body 
Hamiltonian, with $\gamma = -1$, $N = 1000$ and $L = 5$. Solid curve: The analytic 
expression, Eq.\,(\ref{fk22}). The difference between the dots and the curve is hardly 
visible.}
\end{figure}

\begin{figure}
\includegraphics[width=8cm,height=7cm,angle=-0]{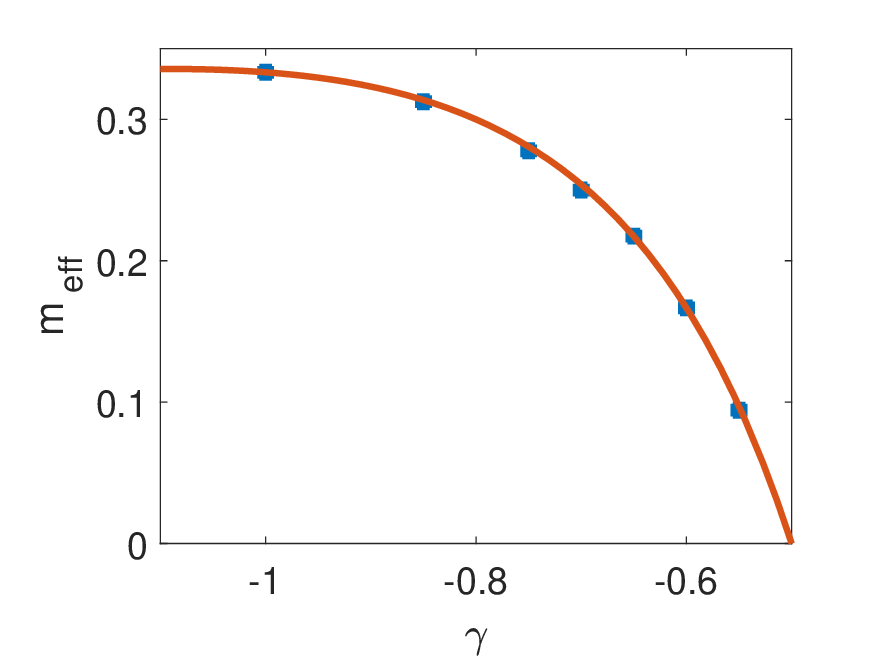}
\caption{The effective mass $m_{\rm eff}$ (in units of $2 M$) from Eq.\,(\ref{effmass}) 
(solid curve), along with the values that we extract from the diagonalization of the 
many-body Hamiltonian (squares), as function of $\gamma$. Here we considered $N = 1000$ 
atoms in the diagonalization of the Hamiltonian.} 
\end{figure}

The eigenvalue equation has the form
\begin{eqnarray}
  H_{k, k-1} f_{k-1} + H_{k, k} f_k + H_{k, k+1} f_{k+1} = E f_k,
  \label{eigeq}
\end{eqnarray} 
where $E$ is the eigenvalue. Assuming that the amplitudes $f_k$ are smooth and 
differentiable functions, we expand \cite{JKMR}
\begin{eqnarray}
  f_{k \pm 1} = f_k \pm \partial_k f_k + \frac 1 2 \partial_k^2 f_k.
\end{eqnarray}
The eigenvalue equation then becomes (to leading order in $N$)
\begin{eqnarray}
\frac 1 2 (H_{k, k-1} + H_{k, k+1}) \partial_k^2 f_k + V(k) f_k = E f_k,
\label{fkk}
\end{eqnarray}
where $V(k) = (H_{k,k-1} + H_{k,k} + H_{k, k+1})$. The above equation may also be written as
\begin{eqnarray}
- \frac 1 {2 \mu}  \partial_k^2 f_k + V(k) f_k = E f_k,
\label{effequ}
\end{eqnarray}
where 
\begin{equation}
  \frac 1 {\mu} = - (H_{k, k-1} + H_{k, k+1}).
\label{muu}
\end{equation}
Therefore, the problem takes the form of a harmonic oscillator. 

The effective potential becomes
\begin{eqnarray}
 V(k) = L + \frac {\gamma} 2 \left(N-1 + 2 L - 2 \frac {L^2} N \right) + 
 \nonumber \\
 + 2 (\gamma+1) k - 3 \gamma k \frac L N
 - 3 \gamma \frac {k^2} N + 
 \nonumber \\
 + 2 \gamma N \left(1 - \frac L N - 2 \frac k N \right) \sqrt{\frac k N \frac {k+L} N}.
\label{effective}
\end{eqnarray}
We observe that $V(k)/N$ coincides with Eq.\,(\ref{emf}), i.e., with the energy of the
mean-field solution, with the substitution $|d_{-1}|^2 = k/N$, as expected \cite{JKMR}. 

Let us now assume that $L \ll k_0$ (we return to this assumption below). Then, 
Eq.\,(\ref{effective}) may be written as
\begin{eqnarray}
  \frac {V(k)} N = \frac {\gamma} 2 + \frac L N (1 + 2 \gamma) 
  - \frac 3 2 \gamma \left( \frac L N \right)^2 +
  \nonumber \\
  + \left[ 2 (1 + 2 \gamma) - 7 \gamma \frac L N \right] \left(\frac k N \right) 
\nonumber \\
  - 7 \gamma \left( \frac k N \right)^2
  - \frac {\gamma} {4 (k/N)} \left( \frac L N \right)^2.
\label{emff}
\end{eqnarray}
Minimizing the above expression with respect to $k$, we find that the minimum
occurs for some $k_0$,
\begin{eqnarray}
  \frac {k_0} N = \frac {1 + 2 \gamma} {7 \gamma} - \frac L {2N} 
  + \frac 7 8 \frac {\gamma^2} {(1 + 2 \gamma)^2} \left( \frac L N \right)^2.
\label{k0}
\end{eqnarray}
The corresponding minimized energy per atom is
\begin{eqnarray}
  \frac {V_{\rm min}} N = \frac {\gamma} 2 + \frac {(1 + 2 \gamma)^2} {7 \gamma}
  + \frac {\gamma} 4 \, \frac {1 - 5 \gamma} {1 + 2 \gamma} \left( \frac L N \right)^2.
\end{eqnarray}
Also, the effective potential $V$ of Eq.\,(\ref{effective}) may be written as
\begin{equation}
  V(k) = {V_{\rm min}} + \frac 1 2 \lambda (k - k_0)^2,
\end{equation}
where 
\begin{equation} 
 \lambda = - \frac {14 \gamma} N. 
 \label{kappa}
\end{equation} 
Finally, $\mu$ from Eq.\,(\ref{muu}) is given by
\begin{eqnarray}
  \frac 1 {\mu} = - \frac 2 {7^2} \frac N {\gamma} (3 \gamma - 2)(1 + 2 \gamma).
\label{meff}
\end{eqnarray}
From Eqs.\,(\ref{kappa}) and (\ref{meff}) one may derive $\omega_{\rm loc}$, 
\begin{equation}
 \omega_{\rm loc} = \sqrt{\frac \lambda {\mu}} 
 = \sqrt{\frac 4 7} \sqrt{(3 \gamma -2) (1 + 2 \gamma)}.
 \label{omm}
\end{equation}
The final expression for the whole excitation spectrum is 
\begin{eqnarray}
   E_n(L) - N e_{\rm loc}(\gamma) = \left( n + \frac 1 2 \right) \omega_{\rm loc}
   + \frac {L^2} N \frac 1 {2 m_{\rm eff}}.
\label{es1}
\end{eqnarray}
From Eq.\,(\ref{es1}) it follows that
\begin{equation}
  \Delta E_n(L) = E_n(L) - E_0(0) = n \omega_{\rm loc} 
  + \frac {L^2} N \frac 1 {2 m_{\rm eff}}.
\label{es11}
\end{equation}
Here $\omega_{\rm loc}$ is given in Eq.\,(\ref{omm}), and $n = 0, 1, 2, \dots$, while 
\begin{equation}
 e_{\rm loc}(\gamma) = \frac {\gamma} 2 + \frac {(1 + 2 \gamma)^2} {7 \gamma},
\end{equation}
and
\begin{equation}
  m_{\rm eff} = \frac 2 {\gamma} \frac {2 \gamma + 1} {1 - 5 \gamma}.
  \label{effmass}
\end{equation}
This equation implies that $m_{\rm eff}$ is a non-monotonic function of $\gamma$,
with its maximum at $\gamma \approx -1.09$. However, for $\gamma \ll -1$ the
truncation to the single-particle states with $m = -1, 0$, and 1 is not expected
to give accurate results. 

In Fig.\,5 we plot the result of Eq.\,(\ref{effmass}) for $m_{\rm eff}$ (solid
curve). The squares in the same figure show the extracted value of $m_{\rm eff}$ 
that results from the numerical diagonalization of the Hamiltonian. More specifically, 
for $N = 1000$ atoms and the values of $\gamma$ which are seen on the plot we 
considered a range of $L$. From the derived dispersion relation $E(L)$ we then 
extracted the curvature and thus $m_{\rm eff}$. We observe that the two results 
agree with each other. 

\begin{figure}
\includegraphics[width=8cm,height=7cm,angle=-0]{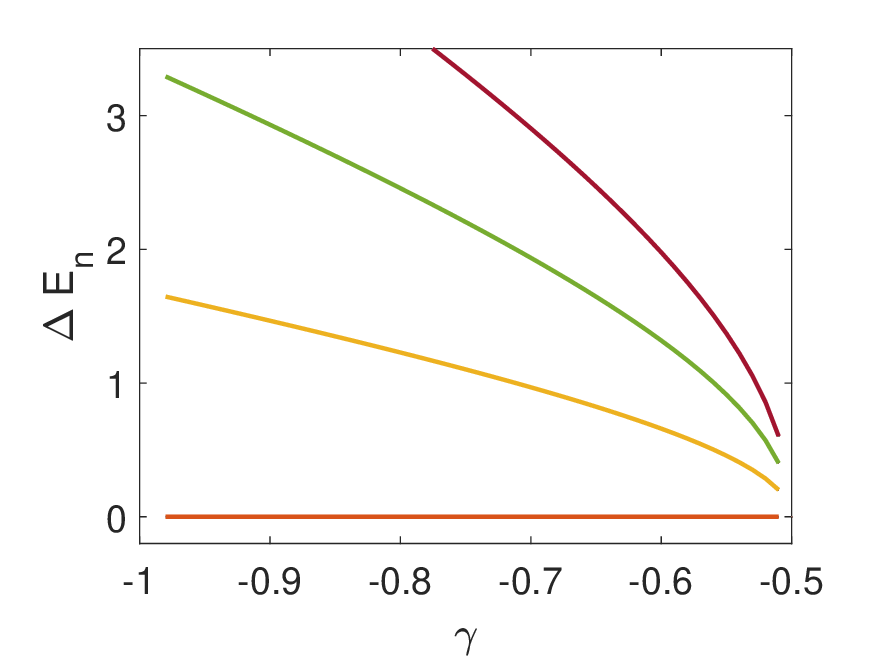}
\includegraphics[width=8cm,height=7cm,angle=-0]{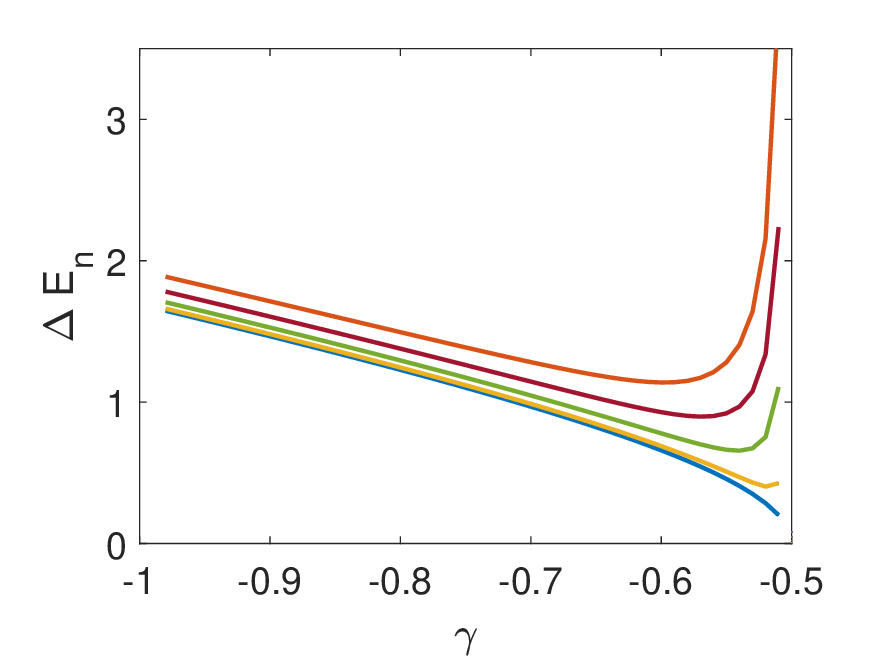}
\includegraphics[width=8cm,height=7cm,angle=-0]{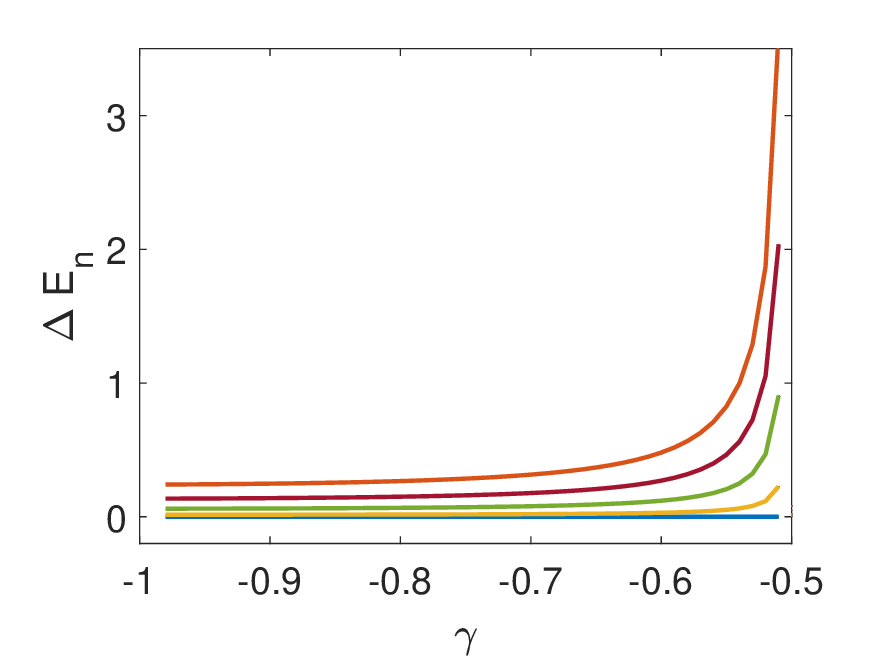}
\caption{The excitation spectrum, $\Delta E_n(L) = E_n(L) - E_0(0)$, Eq.\,(\ref{es11}), 
measured from the energy of the lowest-energy state (in units of $e_1$),  as function 
of $\gamma$, for $N = 100$, in the truncated space of the single-particle states with 
$m = -1, 0$ and 1. Here $-1 \le \gamma \le -0.5$, i.e., the system is in the localized 
phase. In the top figure $n = 0, \dots, 3$ (from the bottom curve to the 
top), and $L = 0$. In the middle figure $n = 0$, and $L = 0, \dots, 4$ (from the bottom 
curve to the top). In the bottom figure, $n = 1$, and $L = 0, \dots, 4$ (from the bottom 
curve to the top).}
\end{figure}

As we mentioned earlier, in deriving the above analytic results we assumed that 
$L \ll k_0$. From Eq.\,(\ref{k0}) it follows that this condition is satisfied if 
$\gamma$ is smaller than $\gamma_c$ by an amount which is of order $L/N$. For the 
values of $L$ that we have considered (of order unity), $L/N$ is of order $1/N$.

Again, there are three important observations here in connection with Eq.\,(\ref{es1}): 
(i) There is no linear term in $L$ in Eq.\,(\ref{es1}), but only a quadratic (due 
to the formation of a localized blob). This is in contrast to the previous case of 
Eq.\,(\ref{excsphom}) (for $\gamma > \gamma_c$, i.e., the homogeneous phase), where 
we have a term which is linear in $L$. Again, as in the previous case, this result 
is general and not an effect of the truncation to the single-particles states 
with $m = -1, 0$ and 1. (ii) The energy levels associated with the quantum number 
$n$ are equally spaced and their energy difference is of order $\omega_{\rm loc}$, 
i.e., of order unity, as in the previous case. On the other hand, the energy due 
to the angular momentum is of order $L^2/N$. (iii) The quantum of energy 
$\omega_{\rm loc}$ vanishes for $\gamma = \gamma_c$ (as in the previous case) 
and is a decreasing function of $\gamma$ (contrary to the previous case). 

Regarding the amplitudes $f_k$, from Eq.\,(\ref{fkk}) it follows that
\begin{equation}
  f_k \propto \exp(-\sqrt{\lambda \mu}(k - k_0)^2/2).
  \label{fk22}
\end{equation}
In Fig.\,4 we plot these amplitudes, as well as the ones that we get from the numerical 
diagonalization of the Hamiltonian, for the case $N = 1000$, $L=5$ and $\gamma = -1$. The 
difference between the two curves is hardly visible.
 
\section{Discussion of the results and experimental relevance}

\subsection{Getting some insight into the excitation spectrum -- connection with the Higgs and the Golstone modes}

The two basic results of our study are Eqs.\,(\ref{excsphom}) and (\ref{es1}),
along with the effective mass $m_{\rm eff}$, Eq.\,(\ref{effmass}). Equations
(\ref{excsphom}) and (\ref{es1}) give analytically the ground-state energy and 
the excitation spectrum of the system that we have considered for the two cases 
of $\gamma$ being smaller (i.e., the localized phase, where we have a ``bright" 
solitary wave) and larger than $\gamma_c$ (i.e., the homogeneous phase). These 
statements refer to the mean-field approach, since, as mentioned also earlier, 
within our approach, the single-particle density distribution is always axially 
symmetric, due to the fact that we work with angular-momentum eigenstates.

When $\gamma > \gamma_c$, i.e., in the homogeneous phase, the spectrum that
results from Eq.\,(\ref{excsphom}) is rather easy to analyse. First of all, the 
term $-\gamma L^2/N$ tends to zero for large $N$ and $L$ of order unity. Also, the 
two terms which involve the two quantum numbers $n$ and $L$ appear in the same 
way in the excitation spectrum, i.e., $(2 n + L) \, \omega_{\rm hom}$. 

The other case, $\gamma < \gamma_c$, where we have the bright solitary wave, is
more interesting. Close to the transition, when $\gamma$ approaches $\gamma_c^-$, 
$\Delta E_n(L)$ is dominated by the term $L^2/(2 m_{\rm eff} N)$, see 
Eq.\,(\ref{es11}), since $\omega_{\rm loc}$ tends to zero in this limit. As before,
the term $L^2/(2 m_{\rm eff} N)$ tends to zero for large $N$ and $L$ of order 
unity. On the other hand, the presence of $m_{\rm eff}$ in the denominator makes 
this term more interesting. According to Eq.\,(\ref{effmass}), $m_{\rm eff}$ is a 
decreasing function of $\gamma$ and it vanishes for $\gamma \to \gamma_c^-$. 
As a result, as $\gamma$ approaches $\gamma_c^-$, the term $L^2/(2 m_{\rm eff} N)$ 
increases, for fixed $L$ and $N$ (see the two lower plots in Fig.\,6). 

To get some insight on the effect of the two quantum numbers, $n$ and $L$, on 
the excitation spectrum, we focus in Fig.\,6 on the case $\gamma < \gamma_c$ and 
consider three different cases. More specifically, we consider $N = 100$ atoms, 
working in the truncated space of the single-particle states with $m = -1, 0$ 
and 1. In the top plot we set $L = 0$ and consider $n = 0, 1, 2,$ and 3. The 
quantum number $n$ corresponds to the energy levels of the effective 
harmonic-oscillator potential that we derived in Sec.\,V. As a result, the 
value $n = 0$ corresponds to the ground state and the three values $n = 1, 2$ 
and 3 correspond to the three lowest excited states of the effective harmonic 
potential. In more physical terms and within the mean-field approximation, these 
are the three lowest excited states of the breathing mode of the localized phase 
(where we have a ``bright" solitary wave). On a mean-field level, these breathing 
modes correspond to the amplitude fluctuations of the order parameter in the 
broken-symmetric state. As a result, these modes are analogous to the Higgs mode.
We stress that these modes are evaluated within the diagonalization of the 
Hamiltonian (and not the mean-field approximation), and for a finite number 
of atoms. Therefore, may argue that they are ``analogous" to the Higgs mode.

In the middle plot of Fig.\,6 we set $n = 0$, and $L = 0, \dots, 4$. For $L \neq 0$,
these are low-lying excited states (of order $1/N$), however $\Delta E_n(L)$ increases 
as $\gamma$ approaches $\gamma_c$, due to the the dependence of $m_{\rm eff}$ on $\gamma$,
that we discussed earlier. Since the angular momentum is the generator of 
translation/rotation around the ring, they correspond to translating the localized blob 
around the ring. Thus, a linear superposition of these modes restores the broken symmetry. 

One may argue that these modes -- which are evaluated within the diagonalization of the 
Hamiltonian, and for a finite number of atoms -- are ``analogous" to the Goldstone modes. 
This is because they result from a broken symmetry, they are massless (in the 
thermodynamic limit of large $N$), and finally a linear superposition of them restores 
the symmetry \cite{UN}. 
Finally, in the bottom plot of Fig.\,6 we set $n = 1$, and $L = 0, \dots, 4$, where 
a similar picture emerges for $\Delta E_n(L)$ as the one in the middle plot.

\begin{figure}
\includegraphics[width=8cm,height=7cm,angle=-0]{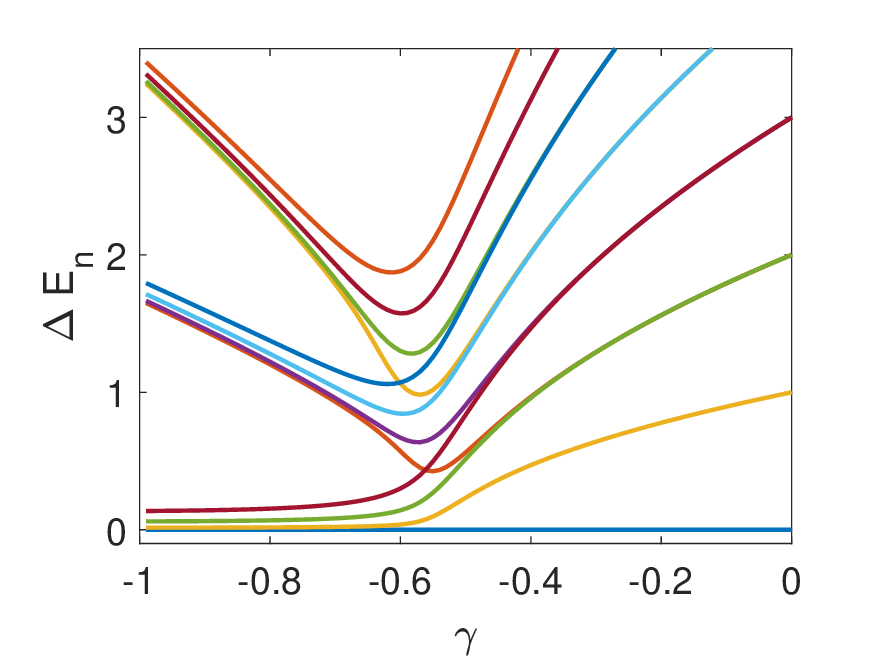}
\includegraphics[width=8cm,height=7cm,angle=-0]{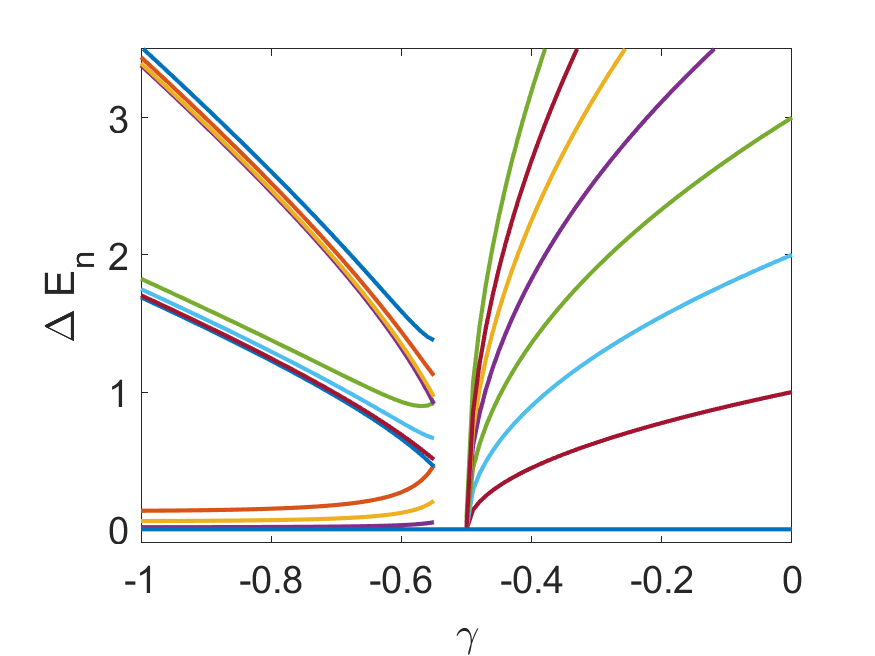}
\caption{The excitation spectrum measured from the energy of the lowest-energy
state (in units of $e_1$), $\Delta E_n(L) = E_n(L) - E_0(0)$, for the first few excited
states, $n = 0, 1, 2$ and $L = 0, 1, 2, 3$, as function of $\gamma$, for $N = 100$, 
in the truncated space of the single-particle states with $m = -1, 0$ and 1. The order
of the curves from the highest to the lowest, on the left part of the plots is: $(n, L) 
= (2,3), (2,2), (2,1), (2,0), (1,3), (1,2), (1,1), (1,0), (0,3), (0,2), \\ (0,1), (0,0)$. 
For $\gamma = 0$ the states with $(n, L) = (2,1)$; $(1,3)$ are degenerate, and also
the pairs with $(n, L) = (2,0)$; $(1,2)$, with $(n, L) = (0,3)$; $(1,1)$ and finally
with $(n, L) = (1,0)$; $(0,2)$. The upper figure is the one that we evaluate numerically, 
from the diagonalization of the many-body Hamiltonian. The lower figure shows the result 
of Eq.\,(\ref{excsphom1}), for $-0.5 \le \gamma \le 0$, and Eq.\,(\ref{es11}), for 
$-1 \le \gamma \le -0.55$. In the interval $-0.55 < \gamma < -0.5$ the result is not 
shown, since, as explained in the text, Eq.\,(\ref{es11}) is not accurate for values of 
$\gamma$ close to $\gamma_c$ (of order $L/N$).}
\end{figure}

\begin{figure}
\includegraphics[width=8cm,height=7cm,angle=-0]{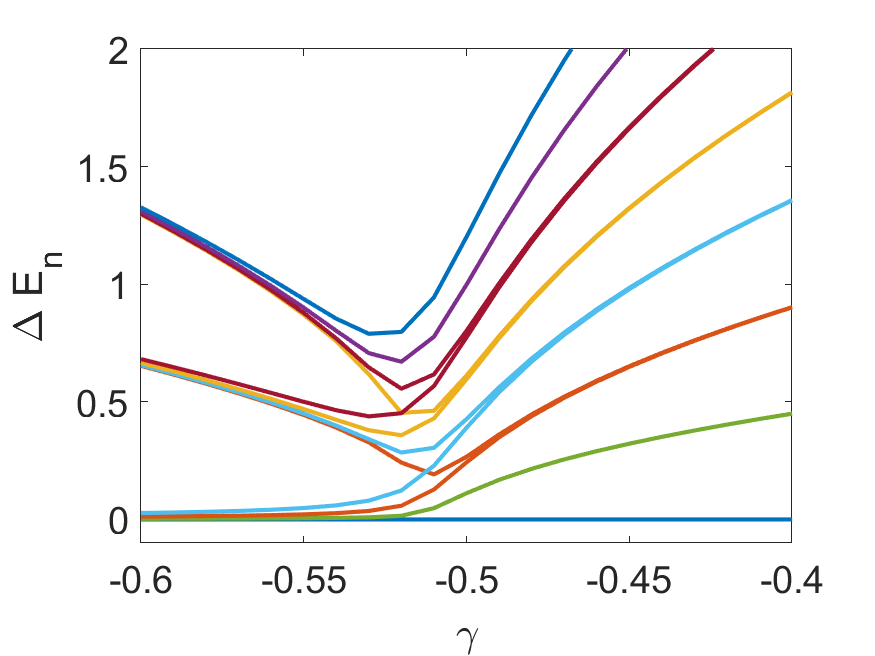}
\includegraphics[width=8cm,height=7cm,angle=-0]{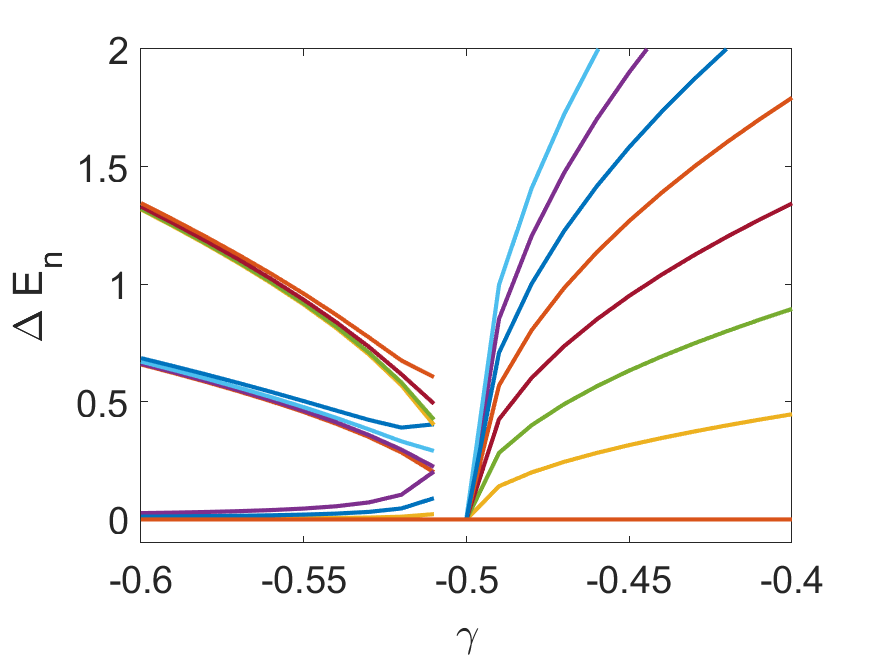}
\caption{The excitation spectrum measured from the energy of the lowest-energy
state (in units of $e_1$), $\Delta E_n(L) = E_n(L) - E_0(0)$, for the first few excited
states, $n = 1, \dots, 7$ and $L = 0, \dots, 3$, as function of $\gamma$, for $N = 1000$, 
in the truncated space of the single-particle states with $m = -1, 0$ and 1. The order
of the curves from the highest to the lowest, on the left part of the plots is: $(n, L) 
= (2,3), (2,2), (2,1), (2,0), (1,3), (1,2), (1,1), (1,0), (0,3), (0,2), \\ (0,1), (0,0)$.
For $\gamma = 0$ the states with $(n, L) = (2,1)$; $(1,3)$ are degenerate, and also
the pairs with $(n, L) = (2,0)$; $(1,2)$, with $(n, L) = (0,3)$; $(1,1)$ and finally
with $(n, L) = (1,0)$; $(0,2)$. The upper figure is the one that we evaluate numerically, 
from the diagonalization of the many-body Hamiltonian. The lower figure shows the result 
of Eq.\,(\ref{excsphom1}), for $-0.5 \le \gamma \le -0.4$, and Eq.\,(\ref{es11}), for 
$-0.6 \le \gamma \le -0.51$. In the interval $-0.51 < \gamma < -0.5$ the result is not 
shown, since, as explained in the text, Eq.\,(\ref{es11}) is not accurate for values of 
$\gamma$ close to $\gamma_c$ (of order $L/N$).}
\end{figure}

\subsection{Comparison between the analytic and the numerical results}

We turn now to the comparison between our analytic results and the ones from the
numerical diagonalization of the many-body Hamiltonian. 

In Figs.\,7 and 8 we plot the excitation spectrum $\Delta E_n(L) = E_n(L) - E_0(0)$ 
as function of $\gamma$, for $N = 100$ and $N = 1000$, respectively, for the first 
few excited states. Such plots were first produced (numerically) in Refs.\,\cite{Ueda0, 
Ueda1, Ueda2}, while similar plots have been published in Ref.\,\cite{droplets}, 
in connection with the formation of a quantum droplet in a ring potential. 

The top plots in the two figures result from the numerical diagonalization of the 
many-body Hamiltonian. The lower plots show what follows from Eq.\,(\ref{excsphom1}) 
(for $-1/2 < \gamma < 0$) and Eq.\,(\ref{es11}) (for $-1 < \gamma < -1/2$). In all 
the cases we have considered, we worked in the truncated space of the single-particle 
states with $m = -1, 0$ and 1. The values of $n$ and $L$ that we have chosen in 
Eqs.\,(\ref{excsphom1}) and (\ref{es11}) are $n = 0, 1$, and 2, and $L = 0, 1, 2$, 
3. Cclearly, the whole excitation spectrum consists of many more states that result 
from higher values of $n$ and $L$. This choice of the values of $n$ and $L$ 
gives rise to the three ``multiplets" of energy levels, which are seen on the left of 
these figures. 

As expected, the agreement between the analytic and the numerical results is better 
for the larger value of $N$. Regarding the dependence of $\Delta E_n(L)$ on $\gamma$, 
for $\gamma = 0$ the excitation energy is 1, 2, \dots, 7, units of $e_1$. 
For $\gamma = 0$, as Eq.\,(\ref{excsphom1}) implies, the states with $(n, L) = (2,1)$; 
$(1,3)$ are degenerate, and also the pairs with $(n, L) = (2,0)$; $(1,2)$, with $(n, L) 
= (0,3)$; $(1,1)$ and finally with $(n, L) = (1,0)$; $(0,2)$. As we argued, for $-1/2 
< \gamma < 0$, $\Delta E_n(L)$ drops as $\gamma$ decreases. 

As $\gamma$ approaches $\gamma_c$, both $\omega_{\rm loc}$ and $\omega_{\rm hom}$ 
tend to zero, i.e., the minimum excitation energy vanishes (for $N \to \infty$), 
or, in other words, there is a ``softening" of the mode. This softening takes
place because for $\gamma = \gamma_c$ the quadratic term in Eq.\,(\ref{emffg}) 
vanishes. In the numerical data from the diagonalization the energy difference 
does not vanish completely because of the finiteness of the system that we have 
considered. 

For $\gamma < -1/2$, on the left half of the plots, we see the excitation spectrum 
of the localized phase, where the Goldstone and the Higgs mode are expected to appear, 
as we argued in the previous subsection. The energy levels for each value of the quantum 
number $n$ in Eq.\,(\ref{es1}) differ by an energy of order unity (see, also, the top 
plot of Fig.\,6). In addition, for each value of $n$, there is a family of modes that 
correspond to the rotational degrees of freedom, $L = 1, 2, \dots$ (see, also, the
two lower plots of Fig.\,6). These energy levels are separated from the one with $L = 0$ 
by a small energy difference, which is of order $L^2/N$. 

\subsection{Experimental relevance}

Regarding the experimental relevance of our study, such ring potentials have been 
realized experimentally, see, e.g., Refs.\,\cite{ring1, ring2, ring3, ring4, ring5, 
ring6, ring7, ring8, ring85, ring9, ring10}. If one considers, e.g., the experiment
of Ref.\,\cite{ring85}, where $^{23}{\rm Na}$ atoms were used, the radius $R$ is 
$\approx 19.5$ $\mu$m and thus $e_1/\hbar = \hbar/(2 M R^2) \approx 3.6$ Hz. Given 
that the scattering length is $a \approx 28$ \AA, for $N \approx 4 \times 10^5$ atoms 
and $S = \pi a_1 a_2$, with $a_i = [\hbar/(M \omega_i)]^{1/2}$, i.e., $a_1 \approx 2.42$ 
$\mu$m  and $a_2 \approx 3.83$ $\mu$m, the dimensionless parameter $\gamma = 
2 N a R/S$ has the value $\approx 1500$. Obviously, in the present case the scattering
length $a$ should be tuned to become negative. If e.g., one reduces simultaneously $N$ 
by two orders of magnitude ($N \approx 4 \times 10^3$), with $a$ being of order 
$-1$ \AA, the value of $|\gamma|$ would become of order unity. 

\section{Summary and Conclusions}

The problem that we have considered in this study, namely Bose-Einstein condensed atoms
which are confined in a ring potential, is an ideal system for the study of various 
effects associated with superfluidity. 

For attractive and relatively weak interactions the single-particle density is homogeneous. 
For stronger (and still attractive) interactions, the atoms form a ``bright" solitary wave. 
In this phase we have a realization of the concept of spontaneous symmetry breaking. As a 
result, in addition to the phenomena which are associated with supefluidity, the system 
that we have considered here provides an example of the well-known Goldstone and Higgs 
modes, in the sense that is discussed in the previous sections and also below. 

The excitation spectrum that we have derived both analytically and numerically 
is characterized by two quantum numbers, $n$ (that is associated with the density 
oscillations of the localized blob) and $L$ (the angular momentum). In the case 
of a homogeneous density distribution they play a similar role with respect to the 
excited states [see Eq.\,(\ref{excsphom})] (apart from a small term, which is of 
order $1/N$).
 
On the other hand, in the regime where the density is inhomogeneous and we have the 
effect of spontaneous symmetry breaking, these two quantum numbers give rise to two 
different ``classes" of excited states [see Eq.\,(\ref{es1})]. The 
excited states which correspond to the quantum number $n$ are low-lying excited 
states, with an excitation energy of order unity (for $L$ of order unity). On a 
mean-field level, they correspond to the breathing modes of the localized blob, 
which is the analogue of the Higgs mode. The excited states which correspond to 
the angular momentum $L$ are also low-lying excited states, with an energy 
difference which is of order $1/N$ (for $L$ of order unity). On a mean-field 
level, the lowest ones are the analogue of the Goldstone modes. 

We should stress that the approach of diagonalization of the many-body 
Hamiltonian that we follow does not provide us with any order parameter. Also, the 
corresponding single-particle density distribution is always axially symmetric, since 
we work with angular-momentum eigenstates. Finally, we work with a finite number of 
atoms. All the above remarks imply that one should not argue that these modes coincide
with the Higgs and the Goldstone modes. On the other hand, the modes that we have 
identified in the derived excitation spectrum have all the characteristics of these 
two well-known modes. 

There are various reasons which make the results of the present study interesting. First 
of all, they provide insight into these modes, which are met in various fields in physics. 
Remarkably, the simplicity of the system that we have considered allowed us to derive 
analytically the whole excitation spectrum (with our results being valid for values of 
$L$ of order unity). In addition, the various physical observables that we managed to 
extract analytically (e.g., the effective mass) may be helpful in trying to measure them 
experimentally and confirm the theoretical predictions in a tabletop experiment.


\begin{thebibliography}{55}
\bibitem{absl} In this calculation we have assumed that $L > 0$. If $L < 0$ then one would
have to work with the Fock states $| k \rangle = | (-1)^{k+L}, 0^{N - L - 2k}, (+1)^{k} 
\rangle$, instead. Then, it follows easily that in Eq.\,(\ref{excsphom}) $L$ should be 
replaced by $|L|$. More generally, the symmetry between $L$ and $-L$ implies that the
energy depends on $|L|$.
\end{thebibliography}
\end{document}